\documentclass[12pt,preprint]{aastex}
\newcommand     \eti    {et~al.}
\begin{document}
\title{Infrared Luminosity Function of the Coma Cluster}
\author{Lei\,Bai, 
George\,H.~Rieke, 
Marcia\,J.~Rieke,
Joannah\,L.~Hinz,
Douglas\,M.~Kelly,
and Myra~Blaylock}
\affil{Steward Observatory,
University of Arizona, Tucson, AZ 85721, USA}
\email{bail@as.arizona.edu}
\begin{abstract}
Using mid-IR and optical data, we deduce the total infrared (IR) luminosities of galaxies in the Coma cluster and present their infrared luminosity function (LF). 
The shape of the overall Coma IR LF does not show significant differences from the IR LFs of the general field, which indicates the general independence of global galaxy star formation on environment up to densities $\sim$ 40 times greater than in the field (we cannot test such independence above $L_{ir} \approx 10^{44}~{\rm ergs~s}^{-1}$).
However, a shallower faint end slope and a smaller $L_{ir}^{*}$ are found in the core region (where the densities are still higher) compared to the outskirt region of the cluster, and most of the brightest IR galaxies are found outside of the core region.
The IR LF in the NGC 4839 group region does not show any unique characteristics. 
By integrating the IR LF, we find a total star formation rate in the cluster of about 97.0 $M_{\sun}{\rm yr}^{-1}$.
We also studied the contributions of early- and late-type galaxies to the IR LF.
The late-type galaxies dominate the bright end of the LF, and the early-type galaxies, although only making up a small portion ($\approx$ 15\%) of the total IR emission of the cluster, contribute greatly to the number counts of the LF at $L_{ir} < 10^{43}~{\rm ergs~s}^{-1}$.
\end{abstract}

\section{Introduction}
Galaxy evolution is largely the story of how the masses, morphologies, and patterns of star formation in these objects vary with environment and cosmological epoch. 
Luminosity functions (LFs) are very important statistical tools for studying evolutionary changes in galaxy populations and provide key observational constraints on galaxy evolution.
The infrared (IR) LF, in particular, by tracing the IR emission of dust heated by star forming activities, remains one of the best probes to study the global evolution of the star formation rate (SFR) with environment and redshift. 
All the indicators show that the SFR is a strong function of epoch.
The average SFR has declined by an order of magnitude to the present epoch from a peak near $z\approx 1$ \citep{Lilly96, Madau96, Pablo05, LeFloch05}.

We also expect the SFR to depend on environment. 
In crowded regions, e.g., the cores of dense clusters, galaxies experience many interactions and mergers, which drive the accumulation of mass and redistribute the gas.
The hierarchical galaxy formation models \citep{Somerville99, Cole00} suggest that galaxies in a cluster gradually lose their gas reservoir as they are accreted into the cluster center.
Those environmental effects are expected to result in a systematic modification of the patterns of star formation in clusters.
In fact, a number of studies \citep{Gomez03, Balogh98} have shown such effects, with a lower SFR in clusters compared with the field. 
The infrared LFs of clusters, as a global indicator of SFR, may also carry such an imprint and differ from the IR LFs of field galaxies. 

Since the launch of the {\it Infrared Astronomical Satellite} (IRAS) and the {\it Infrared Space Observatory} (ISO), there have been many studies of the infrared LF of galaxies \citep[e.g.,][]{Rush93, Serjeant01,Takeuchi03}. 
However, these works concentrate on general field surveys and there is no systematic study of IR LFs with epochs devoted to the cluster population. 
Such a study will enable us to disentangle the evolution of the SFR in different environments and help us understand where and why the changes in SFR occurred.

The recently launched $Spitzer$ Space Telescope with the Multiband Imaging Photometer \citep[MIPS,][]{Rieke04}, capable of high resolution, large sky surveys with high sensitivity, provides us the capabilities necessary for thorough studies of the star forming properties of dense galaxy clusters at different redshifts. 
In this paper, the first one of a series studying the IR LF of clusters up to $z \approx 0.8$ using MIPS 24 and 70 \micron\ observations, we present the IR LF of the Coma cluster.
The Coma cluster, as the nearest rich cluster, provides us an excellent chance to study the properties of the IR LF down to a very faint limit.
It also enables us to study the change of IR LF in different regions of the cluster, as well as the contributions to the total IR LF from different types of galaxies. 
This paper will show how the cluster environment shapes the current star formation in this prototypical dense cluster and it will provide a foundation for the future studies of the SFR patterns in other clusters, both nearby and at high redshift. 
In the paper, we use the cosmological parameter set $(h,\Omega_{0},\lambda_{0}) = (0.7,0.3,0.7)$.
We assume a distance modulus of m - M = 35.0 mag for the Coma cluster at z=0.023 \citep{Struble99}.
 
\section{Observations and Data Reduction}
We used MIPS to observe the Coma cluster in medium scan map mode on Jun 6, 2004.
Our map covered a 4 deg$^{2}$ area centered at $\alpha=12^{h}59^{m}27^{s}, \delta=+27\degr51\arcmin53\arcsec$, which included both the cluster core region and the NGC 4839 region. 
Fig.~\ref{f_sky} shows the region of the map. 
The 24 \micron\ and 70 \micron\ reductions were carried out with the MIPS Data Analysis Tool \citep{Gordon05}. 
The total exposure time was about 88 seconds per pixel for the 24 \micron\ observations and about 40 seconds for the 70 \micron\ ones.

\section{Source Detection}

SExtractor was applied to the images to detect sources automatically and to obtain photometry parameters.
First, the images were analyzed and sky background models were built.
Then the images were background-subtracted and filtered with Gaussian functions with the full width at half maximum (FWHM) matching the FWHM of the MIPS 24 \micron\ and 70 \micron\ point spread functions (PSFs).
All the objects with values exceeding a certain threshold of local background were extracted.
We set this detection threshold, relative to the root-mean-square background noise, at 0.65 for the 24 \micron\ image and 2 for the 70 \micron\ image.
After deblending the adjacent objects and `cleaning' the artifacts due to bright objects, SExtractor gave a final catalog of sources.
SExtractor provides several types of magnitude measurements.
We adopted the MAG\_BEST magnitude.
In most of the cases, the MAG\_BEST magnitude is measured in an adaptive aperture of 2.5 times the Kron aperture, but in crowded fields, it is measured in an isophotal aperture and corrected for aperture losses.
In our data, few regions are crowded.

\section{Completeness}
The completeness of source detection affects the LF directly, so it is important to know the detection limit of the observations.
\citet{Papovich04} studied the source detection completeness of MIPS 24 \micron\ images in several different fields.
Among these fields, the Bo\"{o}tes image has almost the same exposure time as the Coma image and the background levels are also similar, with a mean value of 22.7 MJy sr$^{-1}$ for the Bo\"{o}tes field and 33.4 MJy sr$^{-1}$ for the Coma field.
By inserting artificial sources in the images and performing source extraction on them, \citet{Papovich04} found a 80\% completeness flux density limit of 0.27 mJy in the Bo\"{o}tes field.
As a rough approximation, a simple linear scaling with the square root of the background level gives us a detection limit of 80\% completeness at 0.33 mJy for the Coma field.

At 70 \micron, the Coma data were obtained at a lower detector bias than those in Bo\"{o}tes, resulting in an improvement in overall performance. 
Therefore, we ignore the background difference and adopt the completeness limit of about 80 mJy obtained in the Bo\"{o}tes field with the same exposure time \citep{dole04}.

\section{Matching Spectroscopic Surveys of the Coma Cluster with 24 \micron\ Sources}

To study the infrared LF of the Coma cluster galaxies, we used a spectroscopic sample so that cluster membership could be confirmed.
Fortunately many spectroscopic surveys have been carried out in this region.  
Among them, the catalogs from \citet{Bei03} and \citet{Mobasher01} have the largest overlapping area with the MIPS 24 \micron\ observations and also go to fairly deep detection limits.
\citet{Bei03} generated a catalog (hereafter BvdHC) using all known Coma cluster redshifts in a 5.2 deg$^2$ region.
This catalog covers almost the whole region of the MIPS 24 \micron\ image except a few small patches at the edge and has a 93\% completeness down to Sloan $r' = 16.27$ mag.
\citet{Mobasher01} performed spectroscopic observations on two rectangular regions, one at the cluster core and the other near the NGC 4839 group, each $32.5\times50.8$ arcmin$^{2}$.  
The core region is totally covered by MIPS observations, and the region near the NGC 4839 group is partly covered.
Its completeness is about 60\% for the bright galaxies ($R < 17$) and decreases towards the faint end.
The difference between the Sloan $r'$ filter BvdHC used and the Cousins $R$ filter \citet{Mobasher01} used is small.
Comparing the common objects in these two catalogs gives a difference of $R-r'\sim 0.03~{\rm mag}$, so we do not differentiate them and just use $R$ to refer to both of them in this paper. 
Furthermore, we complemented Mobasher's catalog with the BvdHC down to $R = 16.27$ mag, and generated a merged catalog in these regions (hereafter MBC).

We selected all the galaxies from these two catalogs (BvdHC \& MBC) with 4000 ${\rm km~s}^{-1} \le cz \le 10000 ~{\rm km~s}^{-1}$ as cluster members \citep{Colless96} and cross matched them with our 24 \micron\ and 70 \micron\ sources.
Any 24 \micron\ or 70 \micron\ source within 10\arcsec\ of the optical galaxy was identified as the IR emission from this galaxy.
This search radius is about twice as large as the FWHM of the MIPS PSF at 24 \micron\ and half of the FWHM at 70 \micron.
It allows a displacement in projected distance up to 5 kpc between the optical centers of galaxies and the peaks of their IR emission.
About 90\% of the cluster members with 24 \micron\ emission above the completeness limit have a displacement between optical center and IR emission peak smaller than 5\arcsec, i.e., 2.5 kpc.
When multiple identifications occurred, the one with the smallest distance from the optical galaxy was selected.
Less than 2\% of the sources had multiple identifications. 
Therefore, our final sample is not sensitive to the details of matching infrared and optical sources; few cases yield ambiguous associations, and moderate changes in the acceptance radius have little effect on the results.

Among the 498 Coma galaxies in the BvdHC within the MIPS 24 \micron\ field, 217 have 24 \micron\ counterparts.
In the part of the field covered at 70 \micron\ , 58 were detected out of 477 members.
In the MBC, there were 123 galaxies detected at 24 \micron\ out of 333 galaxies and 33 at 70 \micron\ out of 302 galaxies.
The number of galaxies detected in both bands was 56 for the BvdHC and 33 for the MBC.
Although the total number of galaxies detected in the MBC is less than in the BvdHC, the overlapping area with MIPS observations is also smaller: it is about 0.8 deg$^2$ for the MBC and about 3 deg$^2$ for the BvdHC.
Therefore, the number density of galaxies detected in the MBC is still larger than in the BvdHC, consistent with their different detection limits.

\section{IR Luminosity Function}

\subsection{Determination of total IR luminosity}
Since we detected relatively fewer galaxies at 70 \micron\ than at 24 \micron, we based our LF calculations mainly on 24 \micron\ sources.

To obtain the total IR luminosities of galaxies, which relate to the total flux from 8 - 1000 \micron, a single measurement of flux density at 24 \micron\ is not enough.
We need more constraints.
Using a self-consistent modelling of the spectral energy distributions (SEDs) of galaxies over a broad range of wavelength, \citet{Devriendt99} published a sequence of galaxy SEDs with different IR luminosities based on a sample of nearby galaxies.
Their sample includes normal spirals, luminous IR galaxies (LIRGs) and ultraluminous IR galaxies (ULIRGs).
If we assume these SEDs are a complete assembly of representatives of nearby galaxies, the color correlation of these SEDs should be the same as the color correlation of the Coma galaxies.
More specifically, if we know the color correlation between the ratio of IR luminosity ($L_{ir}$) and 24 \micron\ luminosity ($L_{24}$) and the ratio of flux density at 24 \micron\ ($S_{24}$) and R band ($S_{R}$) from the template SEDs, we will know the color correlation of the Coma galaxies as well. 
Therefore, we can use observational data regarding $L_{24}$, $S_{R}$ and $S_{24}$ to get the total infrared luminosity $L_{ir}$.

However, the assumption that the template SEDs include all the galaxy types in Coma is not correct.
For a cluster as rich as Coma, the early type (E/S0) galaxies dominate the optical emission of the cluster.
In the infrared, the spiral galaxies are generally more luminous than the early type galaxies and hence are the majority of the IR sources.
However, given the sensitivity of MIPS and the closeness of the Coma cluster, we still detected the 24 \micron\ emission from many elliptical galaxies and S0 galaxies.
In the BvdHC, which gives information on the galaxy type, about half of the galaxies detected at 24 \micron\ are early type galaxies, and the rest are mostly spiral galaxies or galaxies without type identification.
The infrared emissions of the early-type galaxies may come from different physical mechanisms or different dust geometry than that of the spirals, and their SEDs may have different shapes and colors than those of the template SEDs.

\begin{figure}
\epsscale{0.8}
\figurenum{2}
\plotone{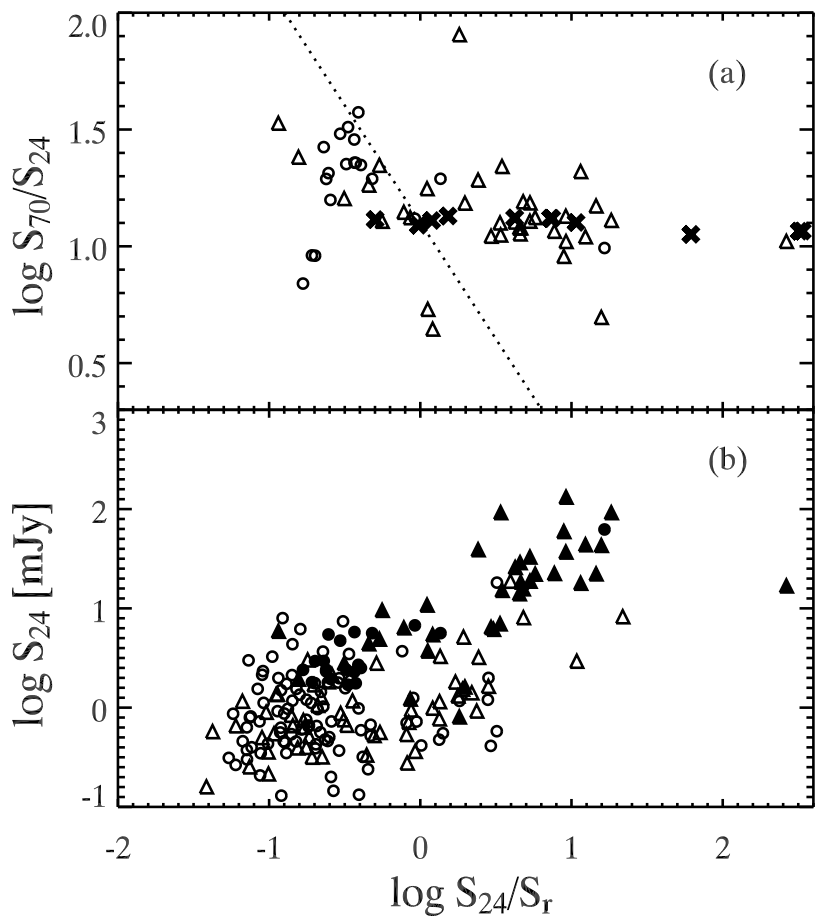}
\caption{The color - color/flux plot of the early- and late- type galaxies.
The open circles and open triangles represent early- and late- type galaxies, respectively.
(a) the 70 - 24 color $vs.$ 24 - R color.
The crosses are the color of the template SEDs.
The dotted line is the detection limit set by the completeness of the 70 \micron\ observations.
(b) the 24 \micron\ flux density $vs.$ 24 - R color.
The galaxies also detected at 70 \micron\ are plotted as filled symbols.}
\label{f_color}
\end{figure}

To check for possible differences between the early-type galaxies and the spiral galaxies in the colors that are crucial to determine the total infrared luminosity, we plot in Fig.~\ref{f_color} the color correlation of the different types of galaxies in the BvdHC.
Panel $a$ in Fig.~\ref{f_color} is the color-color plot of $S_{70}/S_{24}~ vs.~ S_{24}$/S$_{R}$ for the galaxies detected both at 24 \micron\ and 70 \micron\ and with a morphology identification. 
The plot shows that early type galaxies (open circles) have smaller $S_{24}/S_{R}$ ratios on average than the spiral galaxies (open triangles), but their $S_{70}/S_{24}$ ratios are similar to the spiral galaxies. 
Although with a large dispersion, the template SED's (crosses) color correlation represents the average value for the whole galaxy sample fairly well.
If the IR emission of the galaxy mainly comes from dust at a single temperature, then small differences in the $S_{70}/S_{24}$ color indicate a small difference in the $L_{ir}/L_{24}$ ratio for these galaxies.
In panel $b$, we show the 24 \micron\ flux density {\it vs.} $S_{24}/S_{R}$ color of all the galaxies.
The early type galaxies mostly reside in the lower corner of the plot but they are well mixed with the spiral galaxies and show no difference in this correlation from the faint spiral galaxies.
The plot also shows that most of the galaxies with 24 \micron\ flux density larger than 6 mJy are also detected at 70 \micron\ (indicated by the filled symbols).
This result is consistent with the detection completeness at 70 \micron\ assuming the average $S_{70}/S_{24}$ color.
There also appears to be a trend between the 24 \micron\ flux and the S$_{24}$/S$_{R}$ color of the galaxies, which indicates larger IR emission from the redder galaxies. 
Using this trend, we can also obtain a detection limit set by the completeness of the 70 \micron\ observation for panel $a$ shown as the dotted line; the region at the left and below the line is affected by the incompleteness and large uncertainties of the 70 \micron\ measurements. 
From both plots, we find that the early type galaxies are generally less luminous at 24 \micron\ than the spiral galaxies and therefore have lower S$_{24}$/S$_{R}$ ratios, but their color correlation does not differ from that of the spiral galaxies with similar 24 \micron\ flux density. 
This justifies the use of the template SEDs as a complete assembly of all types of galaxies to deduce the total IR luminosity from the $L_{ir}/L_{24}~ vs.~ S_{24}/S_{R}$ correlation.

\begin{figure}
\epsscale{1.0}
\figurenum{3}
\plotone{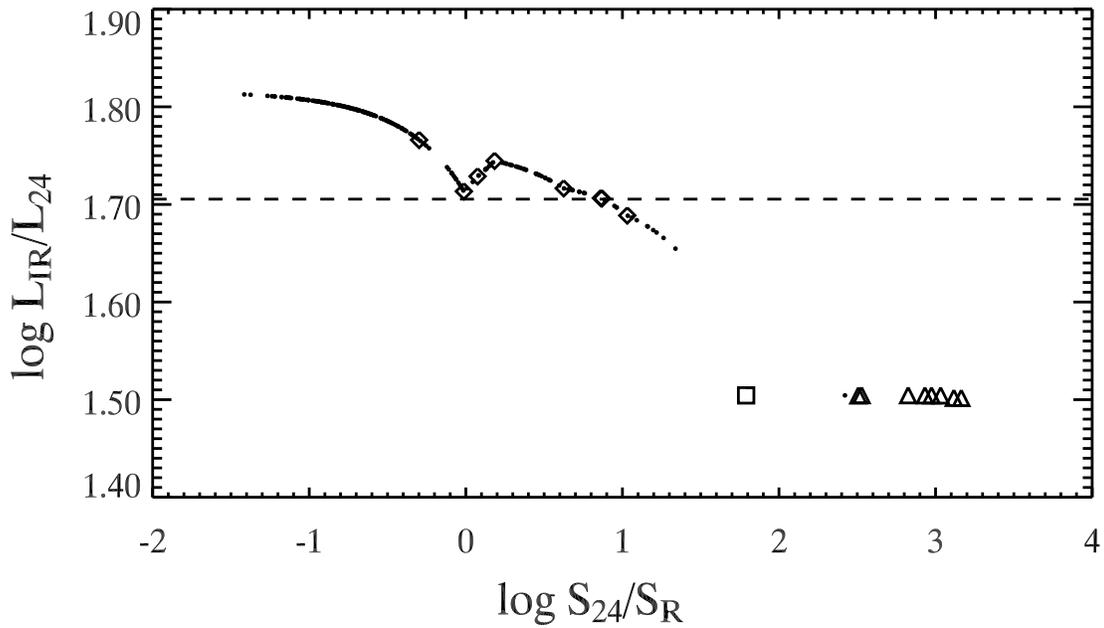}
\caption{The $L_{ir}/L_{24}$ ratio as a function of 24 - R color. The diamonds 
denote the normal spirals, the square a LIRG and the triangles designate ULIRGs.
The dots are the results of the interpolation of the Coma galaxies from their 24 - R colors.
The $L_{ir}/L_{24}$ ratio given by \citep{Lagache03} for a normal spiral is shown as the dashed line.}
\label{f_ratio}
\end{figure}

To obtain the flux densities of the SEDs at different bands, we convolved the SEDs with the response functions of the filters. 
For the 24 \micron\ and 70 \micron\ bands, we also account for the color corrections as described in the MIPS Data Handbook. 
In Fig.~\ref{f_ratio}, we plot the correlation of $L_{ir}/L_{24}~ vs.~ S_{24}/S_{R}$ obtained from template SEDs as well as the value interpolated from the Coma galaxies. 
The log($S_{24}/S_{R}$) ratios of the template spirals range from about $-0.5$ to 1 and those of the LIRGs and ULIRGs from about 2 to 3.5.
Most of the Coma galaxies have log($S_{24}/S_{R}$) ratios smaller than 1.5 and only one source has a color similar to the LIRGs/ULIRGs.
The log($L_{ir}/L_{24}$) ratios of the LIRGs/ULIRGs are almost a constant of $\sim 1.5$.
These ratios for the normal spirals increase slowly with the decrease of $S_{24}/S_{R}$ ratios, with a little (and insignificant) dip at about log($S_{24}/S_{R}) \approx 0$.
Although a simple interpolation onto the correlation works for many of the Coma galaxies, our template SEDs do not cover the range with log($S_{24}/S_{R}) < -0.5$ as the data do, so the color correlation at this end is an extrapolation from the last few points of the template SEDs. 
This unbounded extrapolation may cause systematic errors when deducing the $L_{ir}$ from the $L_{ir}/L_{24}$ ratio.
However, we can find some support for the higher value of the $L_{ir}/L_{24}$ ratio at this end from panel $a$ of Fig.~\ref{f_color}.
Despite the incompleteness and large uncertainties, the panel shows a slightly higher value of the $S_{70}/S_{24}$ ratio than the ratio given by template SEDs towards the lower end of $S_{24}/S_{R}$ ratio.
It is worth noticing that this extrapolation is also consistent with the general expectations for thermal emission: galaxies with smaller values of $S_{24}/S_{R}$ have relatively less emission by warm dust and therefore a higher value of $L_{ir}/L_{24}$.
In the future, a template SED with lower $S_{24}/S_{R}$ ratios will be needed to further constrain the $L_{ir}/L_{24}$ color at this end, but for now we rely on the simple extrapolation. 
With the $L_{ir}/L_{24}$ ratio of each galaxy in hand, we can directly deduce the $L_{ir}$ of each galaxy from its $L_{24}$.

An important result from Fig.~\ref{f_ratio} is that $S_{24}/S_{R}$ is nearly independent of $L_{ir}/L_{24}$ over the luminosity range of interest to us. 
Therefore, our initial galaxy selection on the basis of a visible spectroscopic study will not introduce biases in the infrared properties of the sample. 

To test the method we used here to determine $L_{ir}$, we compare our result with the IR luminosities obtained with the method used by \citet{LeFloch05}.
They adopt a different set of SED templates that are luminosity dependent, e.g., the templates given by \citet{Lagache03}.
Lagache et al.'s galaxy templates include separate SEDs for normal galaxies and starburst galaxies.
Their normal galaxies, again, only include spiral galaxies.
As we can see from Fig.~\ref{f_ratio}, most of the galaxies in the Coma cluster are normal galaxies and only one galaxy has a $S_{R}/S_{24}$ color similar to LIRGs/ULIRGs, so we plot Lagache et al.'s correlation between $L_{ir}/L_{24}$ for normal galaxies in Fig.~\ref{f_ratio} as the dashed line.
This correlation agrees well with the $L_{ir}/L_{24}$ ratios given by the template spirals and thus demonstrates the consistency between our method and that of Le Floc'h.

\begin{figure}
\figurenum{4}
\plotone{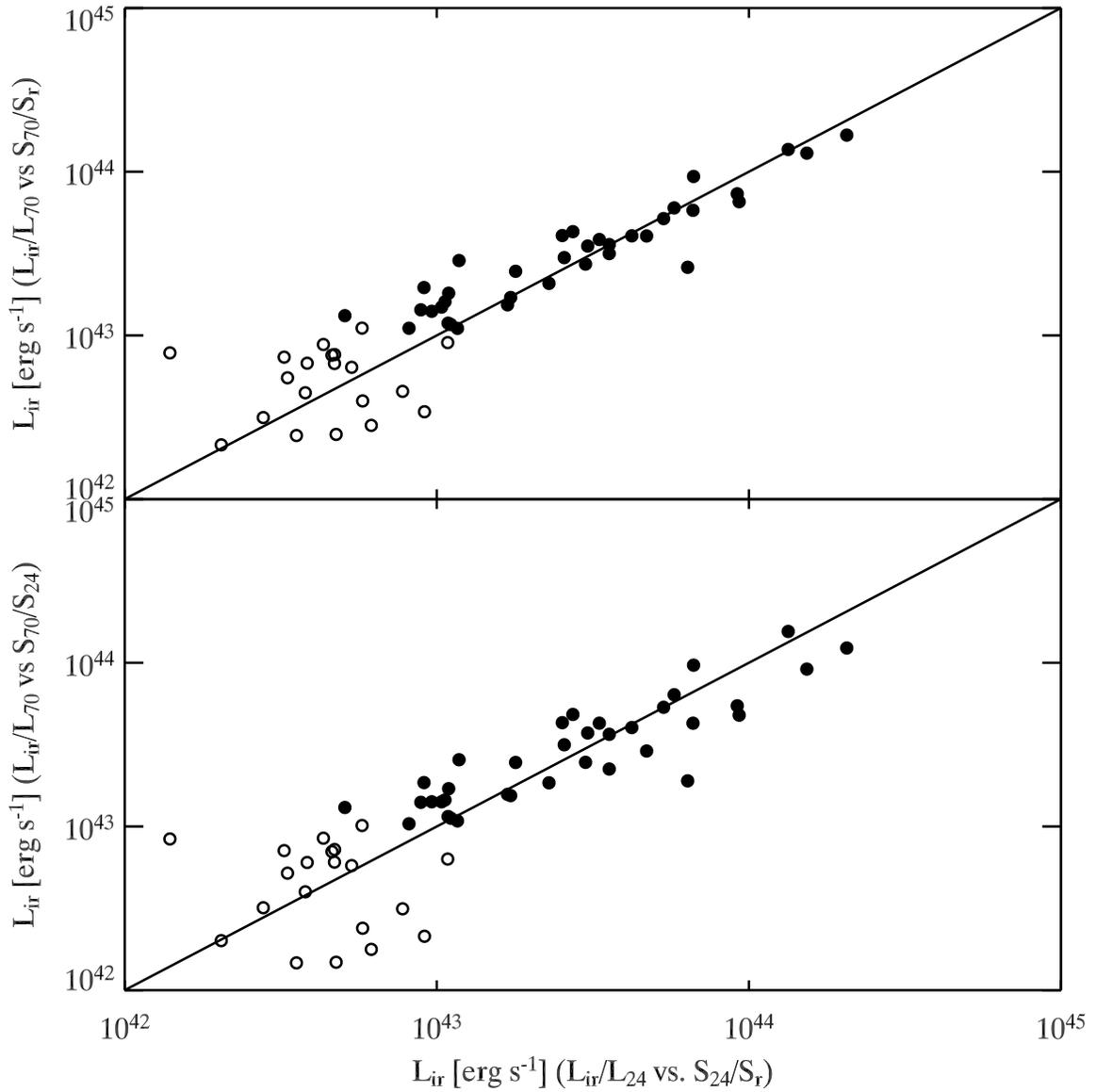}
\caption{Comparisons of $L_{ir}$ obtained from different color correlations. 
The filled circles are the galaxies with $S_{70}$ greater than the 80 mJy completeness 
limit, and the open circles are the galaxies under the limit.}
\label{f_comp}
\end{figure}

To test the self-consistency of our method, we use the data for the members of the BvdHC detected both in the 24 \micron\ and 70 \micron\ bands.
Using the same SED mapping method as before, but, using different combinations of bands, we estimated new values of $L_{ir}$ for these objects.
We plot them against the previous $L_{ir}$ obtained from the correlation between $L_{ir}/L_{24}$ and $S_{24}/S_{R}$  in Fig.~\ref{f_comp}.
The top panel in Fig.~\ref{f_comp} is the $L_{ir}$ obtained from the correlation between $L_{ir}/L_{70}$ and $S_{70}/S_{R}$ compared with the previous one.
The bottom panel is the $L_{ir}$ obtained using the correlation between $L_{ir}/L_{70}$ and $S_{70}/S_{24}$.
The $L_{ir}$ values obtained from different color correlations are generally consistent with a standard deviation of about 0.10 dex for the top panel and 0.16 dex for the bottom panel.
The galaxies with $S_{70}$ under the 80 mJy completeness limit, shown as the open circles in the figure, have large uncertainties in their $S_{70}$ measurement (with errors up to 40 \%) and therefore show a more scattered correlation.
The dispersions of the correlations are generally consistent with the dispersion caused by the uncertainties in the photometric measurements.
It is also possible there are significant contributions to the dispersions by intrinsic color dispersions of the galaxies, as opposed to the tight correlation we assumed in Fig.~\ref{f_ratio}.
Since the dispersions are modest and have zero averages, they will in any case have little effect on the LF we deduce. 

\subsection{Contamination from AGNs}
When we measure the IR emission from galaxies and study their star forming activities, contamination from AGNs is always an issue.
The IR emission of AGNs comes from dust heated by the soft X-ray and ultraviolet emission of the active nuclei rather than from star forming activities; therefore, their SEDs could be very different from the template SEDs we used.

To search for the AGNs in the Coma cluster, we cross matched the {\it Quasars and Active Galactic Nuclei} catalog \citep{Veron03} with the BvdHC and MBC.
The \citet{Veron03} catalog is not complete but it includes almost all the AGNs in the literature.
There are three AGNs detected at 24 \micron\ : D 16, Sy1; NGC 4853, Sy and KUG 1259+280 Sy.
Among them, only NGC 4853 has a $L_{ir}$ greater $10^{43}~{\rm ergs~s}^{-1}$.
These AGNs, so few in number, do not have a noticeable effect on the IR LFs we obtain. 

\subsection{Total IR Luminosity Function}

After testing for the method we used to deduce the total IR luminosity as described above, we calculated the projected IR luminosity function of the Coma cluster.

For the BvdHC, we obtained the number density of galaxies per projected area by directly counting the number of galaxies in each luminosity bin and dividing the number by the projection area.
For the MBC, we assume the completeness function is unity for $R < 16.27$ mag and behaves as described in \citet{Mobasher01} for galaxies fainter than $R = 16.27$ mag.
In calculating the number counts of galaxies, we used the inverse of this completeness function as a weighting factor to correct for the incompleteness.

Both LFs are affected by the completeness of the spectroscopic surveys as well as the IR observations.
The BvdHC spectroscopic completeness is about $R = 16.27$ mag,  and for the MBC, since we already correct for the incompleteness, the completeness limit can be extended to $R = 19$ mag.
The spectroscopic completeness levels of the BvdHC and the MBC are both well above the detection limits in surface brightness for the two surveys \citep{Bei02a,Komiyama02}, suggesting our samples are not limited by the galaxy surface brightness.  
In the end, we need to estimate the IR completeness of samples selected in the optical for the spectroscopic surveys.
We therefore need to relate the two spectral ranges.
About 93\% of our galaxy sample has a $S_{24}/S_{R}$ ratio smaller than 6.5, so we can use this value to set the upper limit of the 24 \micron\ flux density corresponding to the completeness in $R$ band. 
With the linear correlation between $L_{ir}$ and $L_{24}$ given by \citet{Lagache03}, we find the completeness limits of the two spectroscopic surveys in total IR luminosity are $2.6\times 10^{9}~L_{\sun}$ and $2.2\times 10^{8}~L_{\sun}$, respectively.
From section 4, we already know that the 80\% detection limit of the 24 \micron\ observations is about 0.33 mJy, which corresponds to a total IR luminosity of about $1.41\times10^{8}~L_{\sun}$. 
This detection limit is lower than the completeness levels set by spectroscopic surveys.
Thus, the spectroscopic data are the primary limit on the range at which the LFs are free from the effects of incompleteness, with only $\sim$ 7\% incompleteness at the lowest luminosities from the dispersion in IR properties.

In the above calculations of the completeness, we utilize the $S_{24}/S_{R}$ color distribution of our Coma galaxy samples.
However, an issue with these samples is that they are optically selected and therefore they are likely to miss galaxies faint in the optical while bright in the IR.
The IR/optical color distribution of our samples might be tilted towards lower values and we might underestimate the number of galaxies we missed in the calculation of the completeness limit.
To check the color bias of our optically selected samples, we use the catalog of \citet{Karachentsev04}, which provides a nearly complete listing of local galaxies within 10 Mpc. 
The catalog gives the $B$-band magnitude of each galaxy.
We obtained the IRAS data for these galaxies, and therefore have a complete sample which is not constrained at the levels of interest by the optical and/or IR detection limits.
We calculated the IR/optical colors $S_{24}/S_{B}$ of the normal galaxies (not dominated by AGN and below the LIRG luminosity range) in the sample and compared them with those of the BvdHC.
In $B$ band, the BvdHC is complete down to $B=17.5$ mag, and the IR completeness limit of $2.6\times 10^{9}~L_{\sun}$ corresponds to a $S_{24}/S_{B}$ color of $\sim 14.6$.
About 95$\pm9$\% of galaxies in BvdHC have $S_{24}/S_{B}$ color smaller than this value, while in the Karachentsev's catalog, this ratio is only slightly lower, 89$\pm14$\%, among the galaxies with same $B$ band magnitude cutoff. 
That is, the small effect of the dispersion in infrared properties on the overall completeness is confirmed by the behavior of the complete sample of local galaxies.

\begin{figure}
\figurenum{5}
%\plotone{f5.eps}
\plottwo{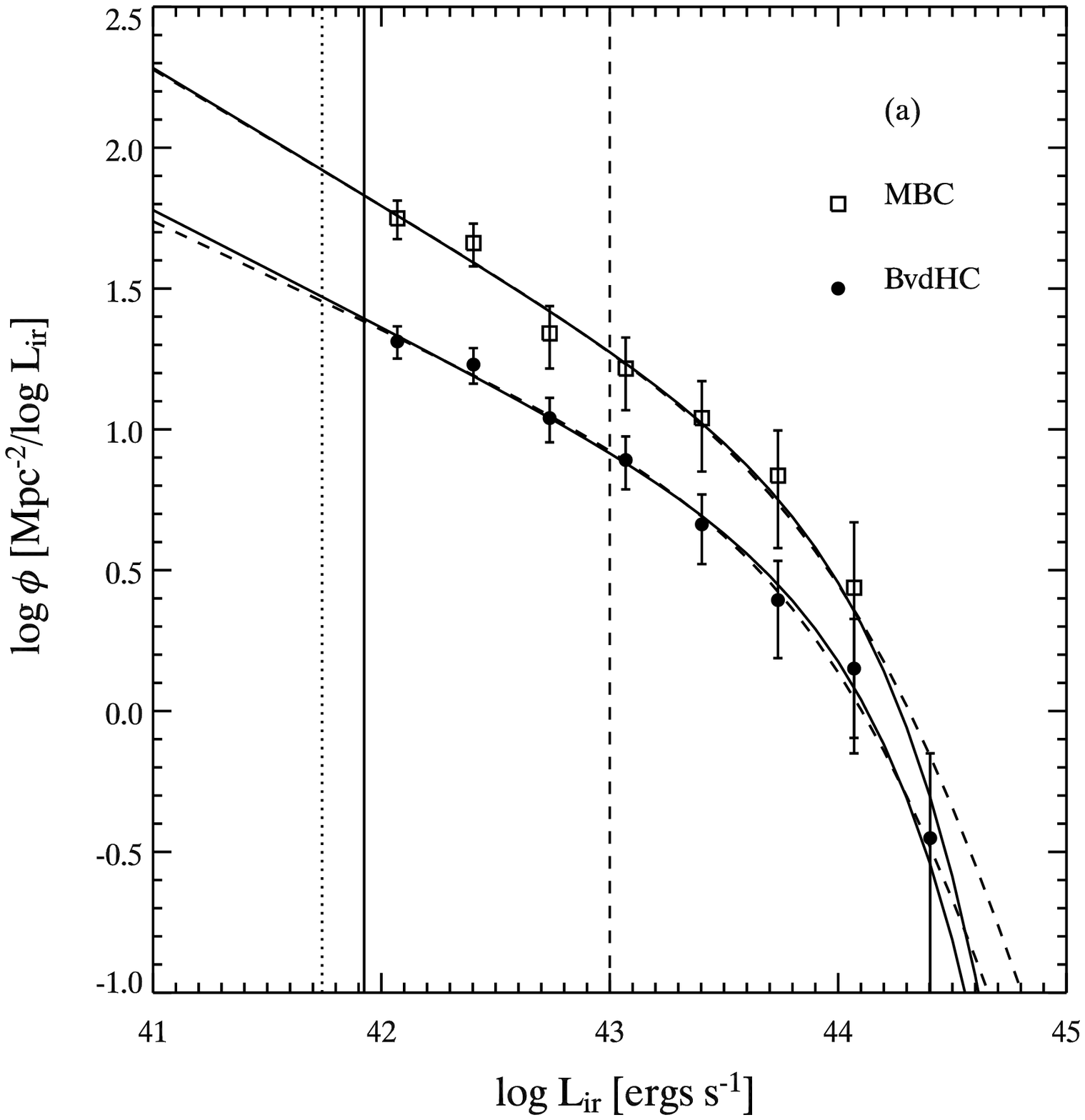}{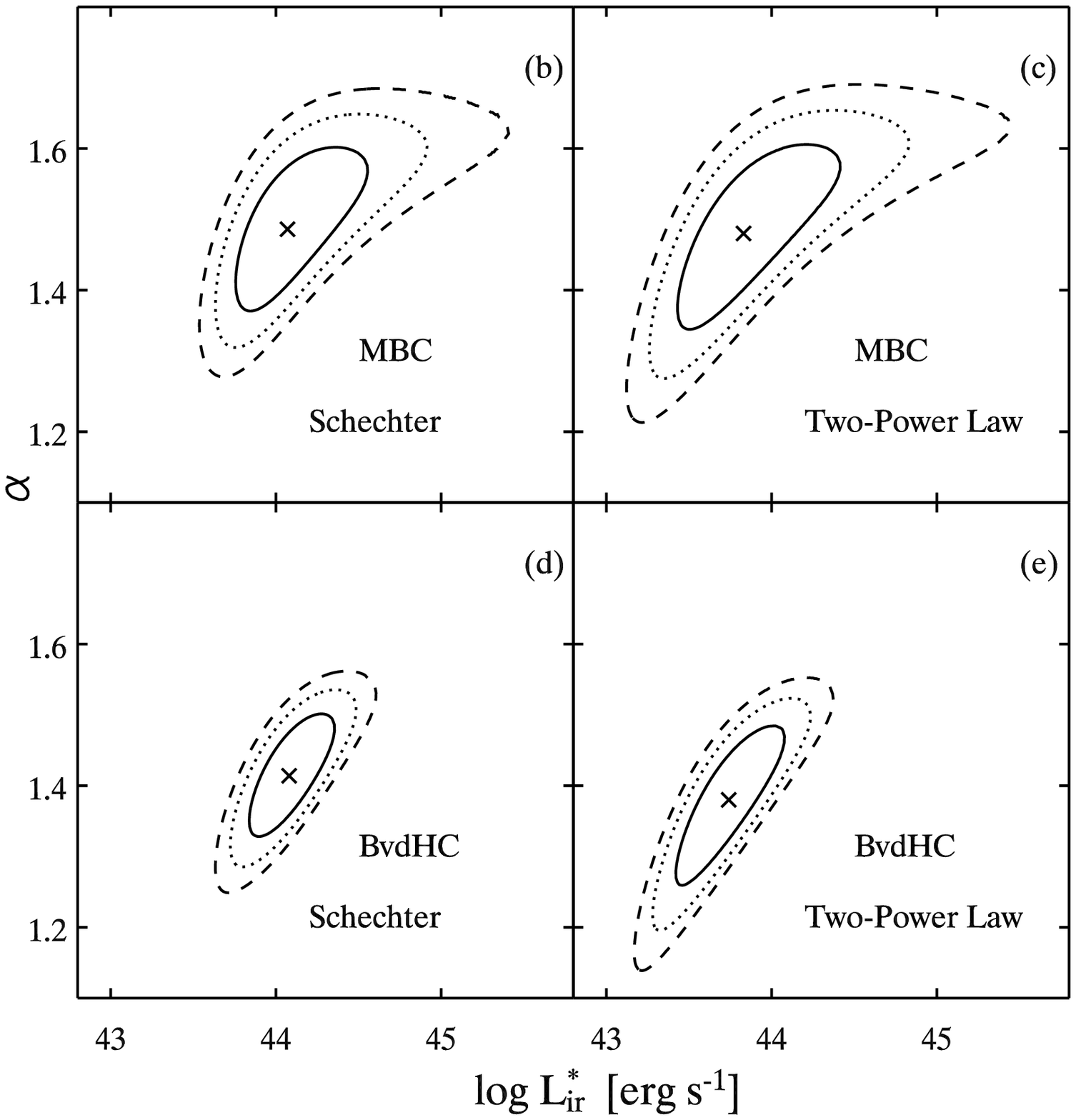}
\caption{(a) The IR luminosity function of the Coma galaxies. The filled circles 
and the open squares represent the LF from the BvdHC and the MBC, respectively.
The error bars denote the statistical error. 
The three vertical lines are completeness limits: the dotted vertical line 
shows the completeness of the 24 \micron\ detections, the solid vertical line 
shows the completeness of the MBC after correcting so far as possible for 
incompleteness, and the dashed vertical line shows the completeness of the BvdHC.
The solid curves are the results of fitting the LF with the Schechter function and the 
dashed curves are the results of fitting with a two-power-law function with a fixed slope at the bright end.
(b),(c),(d) and (e) are the error contours for the fitting parameters $L_{ir}^{*}$ and $\alpha$. The contour levels are 1, 2 and 3 $\sigma$. The best-fitting parameters are indicated by the cross symbols. }
\label{f_LF}
\end{figure}

In Fig.~\ref{f_LF}, we plot the LFs obtained from the two spectroscopic catalogs along with the completeness limits.
The filled circles are results from the BvdHC and the open squares from the MBC.
The dotted vertical line is the detection limit at 24 \micron, the solid line is the limit of the MBC, and the dashed one is the limit of the BvdHC.
The shapes of these two LFs are similar, but the LF from the MBC has an overall higher number density than that from the BvdHC.
The reason for this difference is that Mobasher's spectroscopic survey covered the whole cluster core, where the galaxy number density is the highest, but only a small portion of the outer region, while the BvdHC is based on a much larger area including both the core and the outer region.
Also, the LF from the MBC has a larger variance compared to the LF from the BvdHC because it is based on a smaller sample.
Below the 24 \micron\ detection limit, the loss of faint galaxies due to the limit of the IR observations causes quick drops in both LFs. 
Above this limit, the faint end slope of the MBC LF is steeper than the BvdHC LF, which is consistent with the different completeness limits of these two LFs. 

To have a more quantitative comparison, we fitted these two LFs with some analytical functions.
We discarded all the data points below $10^{42}~{\rm ergs~s}^{-1}$ and use a chi-square minimization method to find the best fitting parameters.
Since we do not have many data points at the bright ends which are critical to determine $L^{*}_{ir}$, it is important to constrain the fitting beyond the last bin for the non-detection of brighter galaxies.
In order to incorporate this factor into the fitting, we calculate the integrated expected galaxy number brighter than the brightest galaxy actually observed for each trial function and use this number to estimate the probability of the non-detection.
This integration is carried out from the brightest luminosity observed to a luminosity 2 orders of magnitude brighter.
The results change little when we extend the integration to a higher upper limit.  
We include the chi-square of this non-detection into the total chi-square value for the minimization process.
We first fitted the LFs with the Schechter function \citep{Schechter76} and the best-fitting parameters we found are:

\begin{equation}
\alpha=1.49^{+0.11}_{-0.11};~ {\rm log}(L^{*}_{ir}/L_{\sun})=10.48^{+0.48}_{-0.31},~{\rm for~ the~ MBC;}
\end{equation}
\begin{equation}
\alpha=1.41^{+0.08}_{-0.08};~ {\rm log}(L^{*}_{ir}/L_{\sun})=10.49^{+0.27}_{-0.24},~{\rm for~ the~ BvdHC.}
\end{equation}
The fitting results are shown as the solid curves in Fig.~\ref{f_LF}.
The $L^{*}_{ir}$ values of these two LFs are very similar.
The parameters of the MBC LF have larger uncertainties due to the few data points at the bright end to constrain the fitting.
It also has a steeper faint end slope than the LF for the BvdHC, which is in agreement with the fact that the MBC has been corrected for incompleteness at the faint end while the BvdHC has not. 
Considering this factor, we expect that the IR LF for the MBC gives a better estimate of the faint end slope than the IR LF for the BvdHC.

A recent work of \citet{Pablo05} studies the 12 \micron\ LF from the 24 \micron\ emission of galaxies using the Spitzer data in two deep field surveys, the Chandra Deep Field South and the Hubble Deep Field North.
Their results, coming from galaxies in the general field, provide a good comparison to our LF in a dense cluster. 
Their LF for galaxies with $ 0.0 < z < 0.2$ gives the Schechter parameter of $\alpha = 1.23\pm0.07$ and log$(L^{*}_{12}/L_{\sun})=9.61\pm0.14$.
With a relation between $L_{ir}$ and $L_{12}$ given by \citet{Takeuchi05}, ${\rm log}L_{ir} = 1.02 +0.972~{\rm log}L_{12}$, the $L^{*}_{12}$ obtained by \citet{Pablo05} corresponds to a total IR luminosity of ${\rm log}(L^{*}_{ir}/L_{\sun})=10.36\pm0.14$, which is only slightly smaller than the value we got; the difference is well within the one sigma error.
However, we found a somewhat steeper slope at the low luminosity end (at about 2$\sigma$ significance).
\citet{Pablo05} suggest that incompleteness may have reduced the value of $\alpha$ in their LF.

\citet{Rush93} obtained a LF using an all-sky 12 \micron\ survey from the {\it IRAS Faint Source Catalog, Version 2}, and fitted it with a two-power-law function
\[\Phi (L) = CL^{1-\alpha}(1+\frac{L}{L^{*}\beta})^{-\beta}.\]
For the non-Seyfert subsample (the majority are normal galaxies, and about 5\% are starburst galaxies), they found the best-fitting parameters are $\alpha = 1.7, ~\beta = 3.6$, and ${\rm log}(L^{*}_{12}/L_{\sun}) = 9.8$.
The average redshift of their non-Seyfert subsample is 0.013, comparable to our sample's average redshift.
Their result has a much steeper slope at the faint end compared to the \citet{Pablo05} result and a little larger $L^{*}_{12}$, which corresponds to log$(L^{*}_{ir}/L_{\sun})=10.55$.
To make a fair comparison to their result, we tried to fit our LFs with the same two-power-law function.
However, since the two-power-law function has too many free parameters and our small sample size gives poor constraints at the high luminosity end, a free fitting failed to give a reasonable result.
We fixed the slope index at the high luminosity end to the best-fitting value given by \citet{Rush93} and kept $L^{*}_{ir}$, $\alpha$ and the normalization free.
The fitting gives very similar results to the Schechter fitting; the best-fitting parameters are \begin{equation}
\alpha=1.48^{+0.12}_{-0.13};~{\rm log}(L^{*}_{ir}/L_{\sun})=10.24^{+0.58}_{-0.39};~\beta=3.6 ~(fixed),~{\rm for~ MBC;}
\end{equation}
\begin{equation}
\alpha=1.38^{+0.10}_{-0.12};~{\rm log}(L^{*}_{ir}/L_{\sun})=10.15^{+0.33}_{-0.31};~\beta=3.6 ~(fixed),~{\rm for~ BvdHC.}
\end{equation}
Again, we note that the MBC LF gives a larger value for the faint end slope and the BvdHC LF gives a more reliable estimate for the $L^{*}_{ir}$ value. 
The MBC LF has a faint end slope a little flatter than Rush's LF, and both LFs give $L^{*}_{ir}$ values smaller than Rush's $L^{*}_{ir}$, with significance about one sigma. 

\citet{Takeuchi03} estimated the 60 \micron\ LF of the local galaxies in the Point Source Redshift survey of $IRAS$.
Based on this LF, \citet{LeFloch05} calculate the total IR LF using the 60 \micron\ total-IR IRAS correlation \citep[e.g.,][]{Chary01} and fit the IR LF with a double-exponential function.
The fitting gives a faint end slope index of 1.23, ${\rm log}(L^{*}_{ir}/L_{\sun})=9.25$ and $\sigma=0.72$, where $\sigma$ is the parameter used to adjust the shape of the bright end of the LF \citep[e.g.,][]{LeFloch05}.
To compare to their result, we fitted our LFs with the same double-exponential function.
Since neither of our LFs has enough data points at the bright end, we have to fix $\sigma=0.72$, the value they gave.
By doing this, we found a faint end slope index of $1.52\pm0.13$ for the MBC LF and log$(L^{*}_{ir}/L_{\sun})=9.39\pm0.37$ for the BvdHC LF.
Again, the Coma IR LF has a very similar $L^{*}_{ir}$ to the field LF.
The faint end slope of Takeuchi's LF is similar to that of \citet{Pablo05}, and they both are flatter than the Coma LF.

The comparisons between the Coma IR LFs and the IR LFs from the general field do not show significant variation of the $L^{*}_{ir}$ value in different environments.
However, a difference in the IR luminosity as small as the 0.3 mag (~0.12 dex in luminosity) difference in $M^{*}_{b_{J}}$ of cluster LF and field LF shown by \citet{Pro03}, is beyond the capability of our study.
The faint end slope of the Coma IR LF is steeper than that of \citet{Pablo05} and \citet{Takeuchi03}, but a little shallower than that of \citet{Rush93}.
This comparison, although complicated by completeness issues, does not support a strong dependence of the shape of the IR LF on environment.

Despite the similarity in the shape of the Coma IR LF and the field IR LF, there might be a large portion of IR-inactive galaxies in the cluster compared with the field.  
Assuming a line-of-sight dimension of the Coma cluster of 13 Mpc, the BvdHC IR LF has a $\phi^{*}$ value - the space density at $L^{*}_{ir}$ - about 45 times larger than the $\phi^{*}$ value given by \citet{Pablo05}. 
Because $L^{*}_{ir}$ is far above our detection limits and those of the spectroscopic surveys (using the proportionality we found for local galaxies between optical and IR luminosities), our study should be complete there. 
With the same assumptions, we find that the average density at $L^{*}_{R}$ is 62.9 $\pm$ 15.2 times that in the field \citep{Geller97, Bei02a}. 
That is, the infrared-emitting galaxy density is only slightly less enhanced in the cluster than the optical galaxy density; there are few extra IR-inactive galaxies in the cluster. 
A second approach to this issue is to examine a sample of field galaxies, and see how many would be detected in the infrared using the same selection method as we have used in the Coma Cluster. 
To implement this approach, we have again used the catalog of galaxies within 10 Mpc from \citet{Karachentsev04}. 
We have compiled the IRAS data for galaxies down to $M_B$ = -17.5 mag, the completeness limit of the BvdHC. 
If these galaxies were at the distance of the Coma Cluster, we find that $0.89 \pm 0.12$ would be detected above our 24 \micron\ limit, whereas the portion of infrared-detected galaxies in the BvdHC down to $M_B$ = -17.5 mag is $0.56 \pm 0.05$.
In agreement with our first estimate, there is only a small deficit of IR-active galaxies compared with the behavior in the field.

It is possible that this difference is partly due to the morphology-environment correlation of galaxies because our Coma sample has about 53\% early-type galaxies while Karachentsev's field galaxy sample only has about 17\% early-type galaxies. 
So we divided our sample into early- and late-type subsamples and calculated the portion respectively.
It turns out this portion just slightly increases in the late-type subsample compared with early-type subsample (57\% $vs.$ 55\%).
Therefore, the morphology-environment correlation can not account for the different portion of IR-active galaxies in Coma and the field.

\subsection{Luminosity Function in Different Regions of the Cluster}

\begin{figure}
\epsscale{1.0}
\figurenum{1}
\plotone{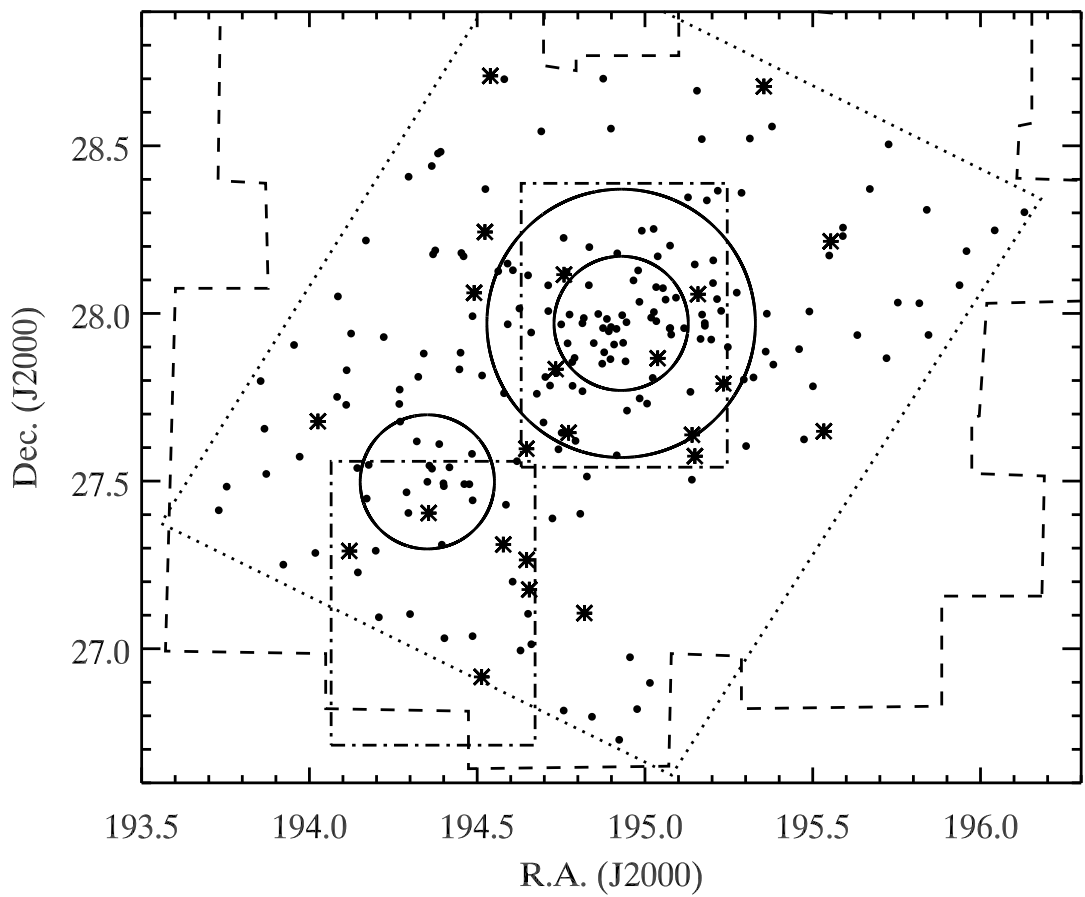}
\caption{The sky map of the Coma cluster. The dots are the galaxy members detected at 24 \micron\ in the BvdHC.
The stars designate the sources with $L_{ir} > 10^{43.3}~{\rm ergs~s}^{-1}$.
The circle in the center defines the Coma core region, with $r < 0.\degr2$, 
and the surrounding annulus is defined as $0.\degr2 < r < 0.\degr4$.
The circular region southwest of the Coma core is centered at NGC 4839 with same radius as the Coma core.
The dotted rectangular region is the MIPS 24 \micron\ coverage.
The dashed region is the survey region of the BvdHC, and the two dash-dotted rectangular regions are of the MBC.}

\label{f_sky}
\end{figure}

Given the large coverage of the BvdHC, we can study the LFs in different regions of the cluster. 
Although the BvdHC is only complete down to $R=16.27$, our previous results show the incompleteness probably will only make the faint end slope a little shallower. 
Since the completeness does not change very much across the cluster, incompleteness will not bias the comparison of LFs in the different regions.
Following \citet{Bei02a}, we define the core region of Coma as the area with $r < 12'$ ($\sim$ 0.3 Mpc) centered at $\alpha=12^{h}59^{m}43^{s}, \delta=+27\degr58\arcmin14\arcsec$. 
We also define an annulus region outside of the core for $12' < r < 24'$ ($\sim$ 0.6 Mpc).
Another interesting area is the group of galaxies around NGC 4839.
It is the second densest region in the cluster and its X-ray emission suggests  that the group is falling into the cluster \citep{Neumann01}.
The interaction between the group and the cluster may trigger star forming activities and therefore affect the IR LF.
Therefore, we also select the circular region centered at NGC 4839 with the same radius as the Coma core region.
We constructed LFs in these three regions. 
Apart from these regions, we took the rest of the area with MIPS coverage as a whole to be the outskirt region of the cluster.
A sky map of these regions and all the galaxy members detected at 24 \micron\ is shown in Fig.~\ref{f_sky}.
There are 40, 56 and 28 galaxies detected in the IR in the Coma core, the surrounding annulus and the NGC 4839 region.
In the outskirt region, 101 galaxies are detected.
The two circular regions have an area of 0.13 deg$^{2}$, the annulus region has an area of 0.38 deg$^{2}$, and the outskirt has an area of about 2.41 deg$^{2}$.
Therefore, the ratios of the projected number density of infrared emitting galaxies in the Coma core, NGC 4839 region, annulus, and outskirt region are about 6:4:3:1. 
Even the lowest-density region has a space density of infrared galaxies roughly 40 times that in the field. 

\begin{figure}
\epsscale{.80}
\figurenum{6}
\plotone{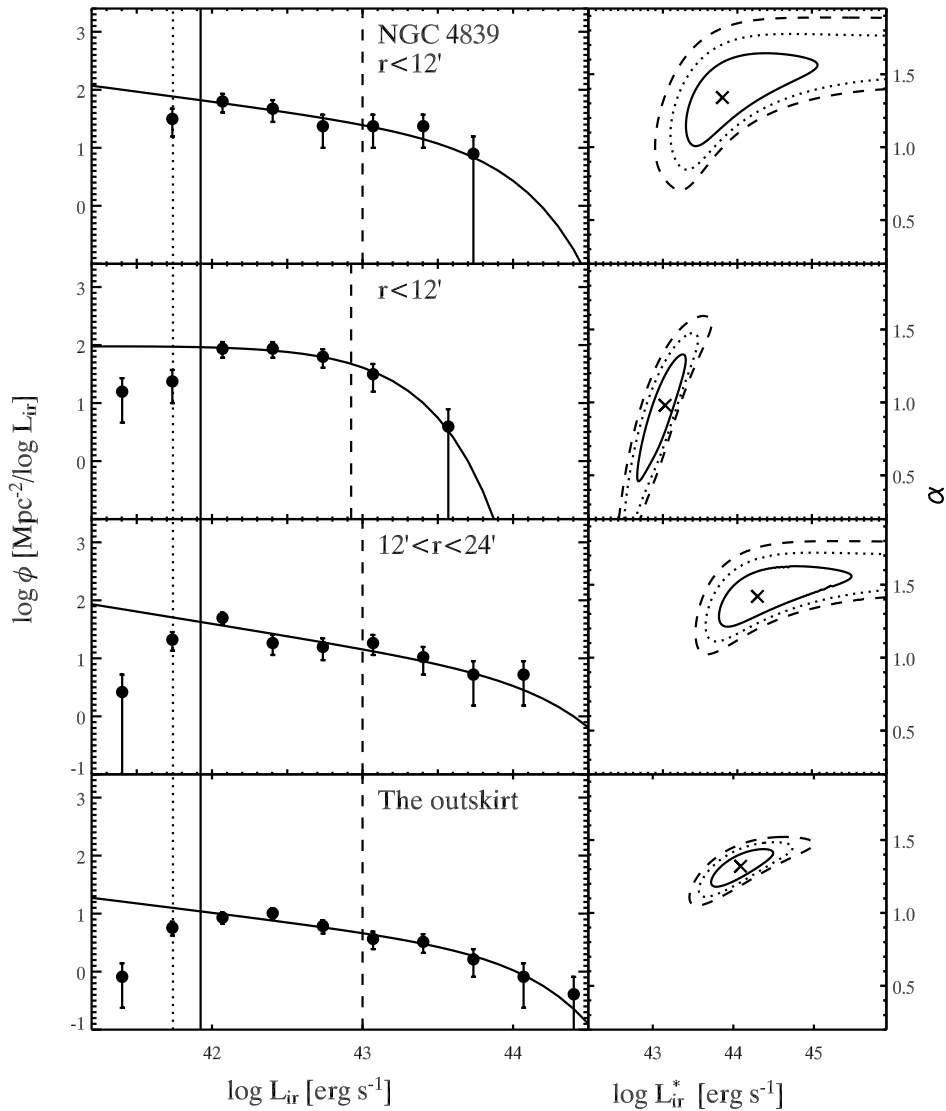}
\caption{The LFs of the galaxies from the BvdHC in different regions of the Coma cluster and the corresponding 1, 2, 3 $\sigma$ error contour maps for the Schechter function fitting.
The vertical lines are the same as in Fig.~\ref{f_LF} and the solid curves are the best fitting Schechter functions. The best-fitting parameters are indicated by the cross symbols on the error contour maps.}
\label{f_region}
\end{figure}

All the LFs were fitted with the Schechter function. 
The results are shown in Fig.~\ref{f_region} and the best fitting parameters are listed in Table 1.
Because the small number statistics in small regions cause large uncertainties in the Schechter function fitting, simple comparisons of the best fitting parameters between these LFs are ambiguous and need to be taken with care.   
The large uncertainties at the bright end of the LFs may cause very different fitting results for the exponential cut-off of the Schechter function and, therefore, unreliable $L_{*}$ values.
However, interesting variations are apparent in other aspects of the LFs. 
The Coma core region has a flatter faint end and fewer luminous galaxies compared to the LFs in the other regions. 
In particular, when we compare it with the LF in the annulus region, which has similar total number counts as in the core, it is apparent that galaxies in the core region are lacking at the high luminosity end. 
A flatter faint end in the core indicates a lack of faint galaxies as well.
The NGC 4839 and the annulus regions have similar number densities and their LFs are not very different from each other at the faint end.
They both have a steeper faint end LF than the Coma core.
At the bright end, it seems that the annulus region has more luminous galaxies than the NGC 4839 region does, but the difference is not significant given the uncertainties.
The LF in the outskirt region is better constrained at both faint and bright ends due to its larger number of galaxies.
Its faint end slope is steeper than the Coma core region but shallower than the annulus and NGC 4839 region.
However, this faint end slope is largely constrained by the lowest point in our fitting process (the point at $L_{ir}=10^{42.05}~{\rm ergs~s}^{-1}$) and we suspect that incompleteness may have a more severe effect on this point in the outskirt region than in the other regions (e.g., because of the lower density of cluster members on the sky).
If we discard this point in our fitting, we have a much steeper faint end slope with $\alpha = 1.52^{+0.16}_{-0.17}$.
With this correction, there appears to be a trend of steeper faint end slope toward the outer regions of the cluster, similar to the behavior in the optical bands \citep{Bei02a, Bei03}.
Although the faint end slope may be questionable, the bright end of the LF in outskirt region is well constrained and has a very similar $L_{*}$ value as the total LF. 
In summary, we found that the Coma core region lacks both very faint and very bright galaxies compared with the outer regions. 
The NGC 4839 region does not show significant difference in the LF from that of the annulus region with similar number density. 
It is also worth noting that all the galaxies with $L_{ir} > 10^{44}~{\rm ergs~s}^{-1}$ reside outside of the core region.

All the LFs in the Coma cluster calculated above are actually the 2-D projection of the real LF, and therefore will be affected by projection effects \citep{Valotto01, Bei02b}.
The projection effect is most serious in the core region and it will probably make the faint end slope steeper. 
That is, the flat faint end slope of the core region may become even flatter after deprojection.
On the other hand, the projection effect will make the lack of bright galaxies in the core region more severe.
Fig.~\ref{f_sky} also shows the galaxies with $L_{ir} > 10^{43.3}~{\rm ergs~s}^{-1}$ as star signs.
They are more or less uniformly distributed in the whole region, without any concentration in the core region or the NGC 4839 region. 
The deprojection from this 2-D distribution will make the bright galaxies move further outward. 

The interpretation of the galaxy population variations across the cluster only from the change of the shape of the LFs may be misleading without knowledge of the fraction of galaxy members detected in the IR.
We already know that the overall fraction of optical galaxy members detected at 24 \micron\ is about 44\% for the BvdHC.
This fraction is smallest in the core region, at about $37\pm7\%$, and it is about $46\pm4\%$ in the outer region.
The difference is not very significant considering the large statistical errors. 
If we only consider the fraction for the galaxies brighter than the completeness of the BvdHC, the difference is even smaller, with the fraction about $54\pm9\%$ in the core and $57\pm5\%$ in the outer region.
Also, a correction for the projection effect, if possible, would make the difference even smaller.
Thus, the fraction of the galaxy members detected in the IR does not change very much across the cluster, providing a uniform foundation for the comparison of the shape of the IR LFs. 

\subsection{Contribution of the Different Types of Galaxies to the Total LF}
\begin{figure}
\figurenum{7}
\plotone{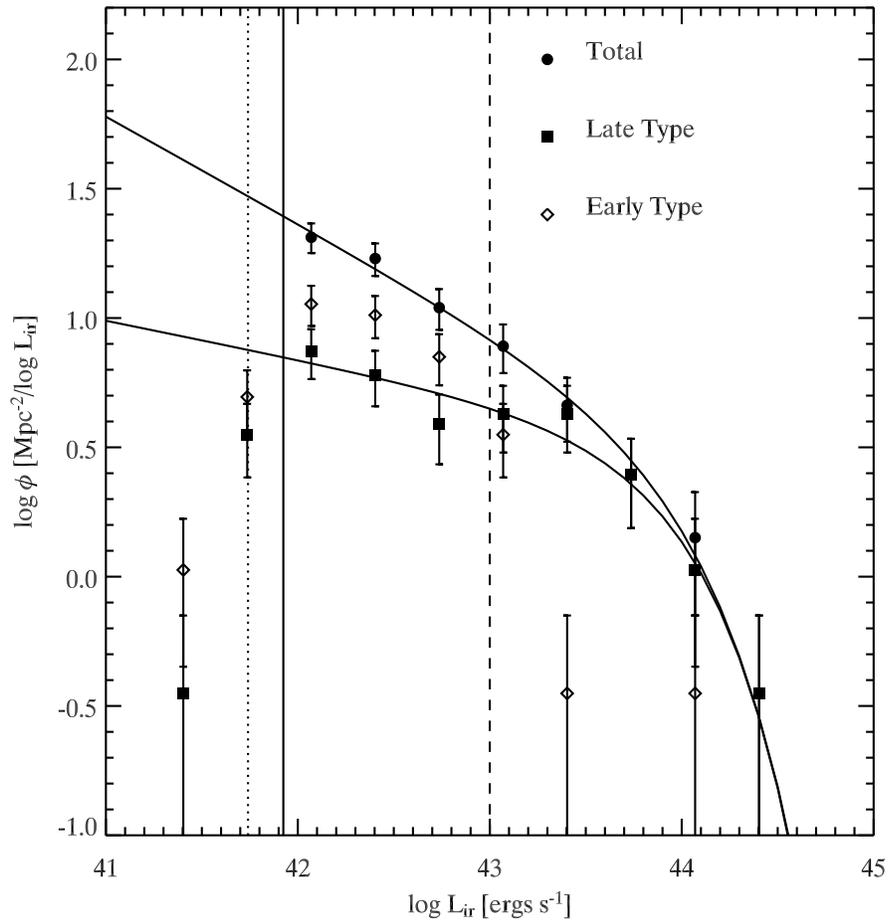}
\caption{The contributions of the early- and late- type galaxies to the LF.
The galaxies are all from the BvdHC.
The vertical lines are the same as in Fig.~\ref{f_LF}.
The total LF and the LF of the late-type galaxies are fitted with the Schechter function, 
and the results are shown as the solid curves.}
\label{f_type}
\end{figure}

Since the BvdHC also has morphology information for each galaxy, we
can study the contribution of early type (E/S0) and late type galaxies to the total LF.
In Fig.~\ref{f_type}, we plot the LF of the late type galaxies and early type galaxies along with the total LF for comparison.
The late type galaxies here include all spirals and irregulars.

From Fig.~\ref{f_type}, we find that the early type galaxies make a larger contribution to the number counts of the LF than the late type galaxies at $L_{ir} < 10^{43}~{\rm ergs~s}^{-1}$, while the late type galaxies dominate the bright end of the total LF.
This behavior indicates that although the late-type galaxies dominate the bright population, there are more faint early-type galaxies than faint late-type galaxies.
However, we note that at the faint end, it is possible to misidentify a spiral as an S0 galaxy and, therefore, the number of early-type galaxies may be overestimated.
We fit the LF of late type galaxies with the Schechter function, and the best-fitting parameters are 
\begin{equation}
\alpha=1.15^{+0.24}_{-0.28};~{\rm log}(L^{*}_{ir}/L_{\sun})=10.44^{+0.50}_{-0.42}
\end{equation}
This LF has a flat faint end and a similar $L^{*}_{ir}$ to the total LF.
Although the $\alpha$ we derive would have a higher value if the incompleteness were taken into account, it is still smaller than the index of the total LF that is affected by the incompleteness in the same way.
The steeply rising faint end of the total LF is boosted by the increasing number of early type galaxies with low IR luminosity. 

Using ISO data, \citet{Pozzi04} deduced the 15 \micron\ LF of the European Large Area ISO survey (ELAIS).
The index of the faint end slope they found for the spiral galaxies with $z < 0.2$ is very close to our value, with $\alpha = 1.10 \pm 0.25$.

Using the LFs of early type and late type galaxies, we can calculate the surface density of the total IR luminosity of these two groups down to the detection limit of the 24 \micron\ observations.
It turns out that the surface density of IR luminosity contributed by early type galaxies is only about 15\% of the total surface density.
Therefore, the early type galaxies make a rather small contribution to the total IR luminosity of the cluster, but they make a significant contribution to the number counts of faint galaxies and therefore affect the shape of the LF. 

\subsection{Measuring the SFR from the IR LF}
\begin{figure}
\figurenum{8}
\plotone{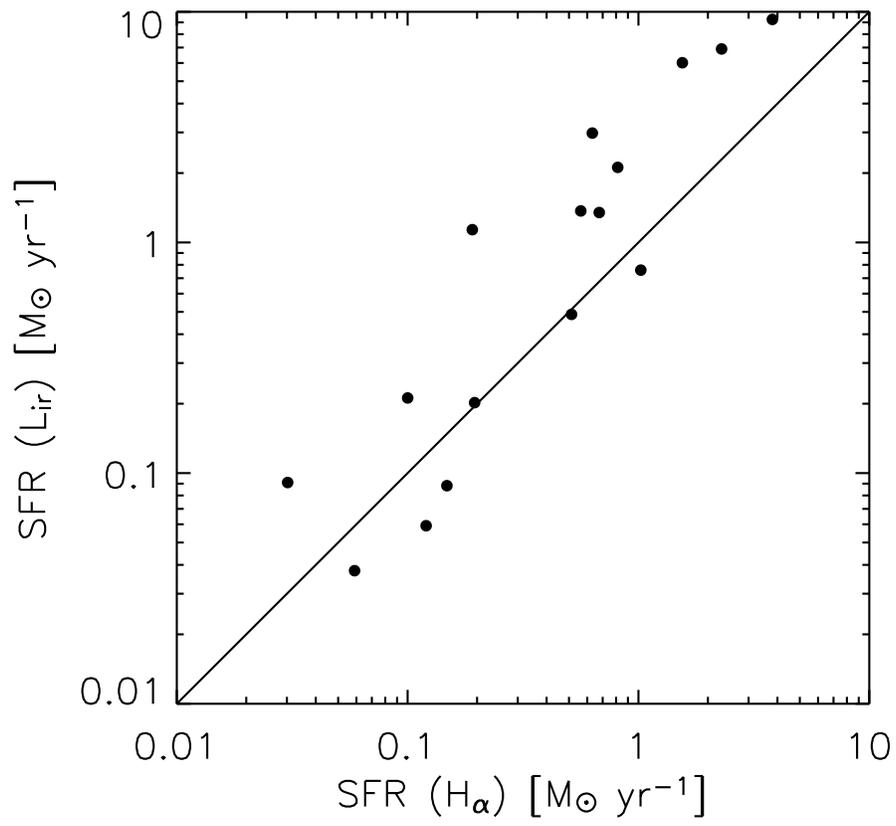}
\caption{Comparison of the SFR measured from $L_{ir}$ and H$\alpha$ luminosity.}
\label{f_SFR}
\end{figure}

Since IR luminosity is a good tracer for star-forming activity, the IR LF allows us to estimate the total SFR of the cluster.
Although this paper is the first work reaching such a depth in the IR luminosity of the Coma galaxies, which means detecting a lower level of star forming activities, there are other works measuring the SFR at a higher level by measuring the ionization lines. 
\citet{Iglesias02} used a deep wide field H$\alpha$ survey of the Coma cluster to deduce the H$\alpha$ LF.
They detected 22 sources in the H$\alpha$ band.
Five of them are not in either the BvdHC or the MBC.
We detected all the rest of them at 24 \micron\ and hence obtained the $L_{ir}$ for them.
Using the conversion formula given by \citet{Rob98}
\begin{equation}
SFR(M_{\sun}{\rm yr}^{-1})=4.5\times10^{-44}L_{ir}({\rm ergs~s}^{-1})
\end{equation}
we deduce the SFR for these objects and compare them with the SFR given by \citet{Iglesias02}.
The result is shown in Fig.~\ref{f_SFR}.
The two results are basically consistent, but the SFRs measured from $L_{ir}$ are larger on average, and this discrepancy is more pronounced in the galaxies with higher SFR.
This discrepancy was also found by \citet{Kennicutt98} when he compared the SFR deduced from the measurement of $L_{ir}$ and the Br$\gamma$ emission line.
He justified the discrepancy by citing effects of extinction and the heating of dust by longer lived stars than those exciting the emission lines.  
These arguments are also applicable to our case.

The H$\alpha$ survey has a smaller area coverage than the MIPS observations and is much less complete than the IR survey.
For the 18 objects detected with both H$\alpha$ and 24 \micron\ emission, 16 have $L_{ir} > 10^{43}~{\rm ergs~s}^{-1}$, but in our sample we have 40 sources with $L_{ir} > 10^{43}~{\rm ergs~s}^{-1}$.
Using the H$\alpha$ LF, \citet{Iglesias02} showed the SFR density of the Coma cluster to be 1.36 $M_{\sun}{\rm yr}^{-1}$Mpc$^{-3}$ by integrating the best fitting function over the whole range of luminosities and assuming the radius of the Coma cluster to be 6.5 Mpc.
In our case, we calculate the total IR luminosities from the best fitting Schechter functions in the range of $10^{42}~{\rm ergs~s}^{-1}<L_{ir}<10^{45}~{\rm ergs~s}^{-1}$.
The upper limit corresponds to the brightest galaxies actually observed and the lower limit excludes the part of the LF with serious incompleteness.
The formula we used to convert the IR luminosity to SFR will also become problematic below this lower limit. 
Fortunately, from the shape of the LF, the galaxies in the $10^{42}~{\rm ergs~s}^{-1}<L_{ir}<10^{45}~{\rm ergs~s}^{-1}$ luminosity range dominate the total luminosity and there is only a small difference between our integration and an integration over the whole luminosity range.  
The total SFRs of the cluster, deduced from the best fitted Schechter functions of the MBC and the BvdHC IR LFs, are 24.48 $M_{\sun}{\rm yr}^{-1}$Mpc$^{-2}$ and 11.41 $M_{\sun}{\rm yr}^{-1}$Mpc$^{-2}$ respectively.
The lower value given by the BvdHC LF is because the BvdHC covers more outskirt regions where the IR galaxy densities are relatively smaller.
The BvdHC is also less complete than the MBC, and thus underestimates the contribution from the faint IR galaxies.
The total SFR in the 8.5 Mpc$^{2}$ area of the BvdHC survey and MIPS observation is about 97.0 $M_{\sun}{\rm yr}^{-1}$.

If we assume the region we observed in the IR has the same line-of-sight dimension as assumed by \citet{Iglesias02}, e.g., $\sim$ 13 Mpc, these two IR LFs give SFR densities of about 1.88 $M_{\sun}{\rm yr}^{-1}$Mpc$^{-3}$ and 0.88 $M_{\sun}{\rm yr}^{-1}$Mpc$^{-3}$.
The SFR density deduced from the H$\alpha$ LF is smaller than the value given by MBC but larger than that given by BvdHC.
However, considering the \citet{Iglesias02} survey also covers mostly the central region, we calculate the SFR from the BvdHC in the region of the H$\alpha$ survey coverage and obtain a larger SFR of 1.7 $M_{\sun}{\rm yr}^{-1}$Mpc$^{-3}$.
Therefore, the IR LF gives a more complete estimate of the total SFR of the cluster than the available H$\alpha$ LF. 
If we adopt 0.88 $M_{\sun}{\rm yr}^{-1}$Mpc$^{-3}$ as the general SFR density of the Coma cluster, we find it to be about 60 times larger than the SFR density of the general field \citep{Pablo05}.
This difference is comparable to the difference in $\phi^{*}$ between these two LFs.
Thus, the higher SFR density in the Coma cluster is mainly due to the overall higher IR galaxy density in the cluster than in the field, not to any differences in the shape of the IR LF.

\section{Discussion}
Despite the evidence that the SFRs of galaxies are different in cluster and field regions \citep{Gomez03, Balogh98}, the comparison of our IR LFs of the Coma cluster with the IR LFs of field galaxies does not support a strong dependence of the shape of the IR LF on environment. 
However, the measurements of the SFR in \citet{Gomez03} and \citet{Balogh98} are based on the bright galaxies ($R< 17$ mag) with ionizing emission lines, which approximately corresponds to the galaxies with $L_{ir} > 10^{43}~{\rm ergs~s}^{-1}$.
In fact, \citet{Gomez03} found that the correlation of the SFR and the environment is most noticeable for the strongly star-forming galaxies.
So, it is very possible that our IR LF, because it is not well constrained at the bright end, does not show a difference from the field LF simply because of the lack of enough very luminous IR galaxies in the small sample to draw any meaningful conclusions. 
In any case, the similar $L_{ir}^{*}$ of the Coma cluster to that of the general field LF, as well as the fairly good fit with similar functions up to  $L_{ir} \approx 10^{44}~{\rm ergs~s}^{-1}$, are evidence against a strong correlation between the global SFR for infrared-bright galaxies and their environment.

Explanations for this behavior may come from some recent works by \citet{Balogh04a, Balogh04b}.
\citet{Balogh04a} studied SFRs of galaxies in group and low-density environments and found that although the fraction of the star-forming galaxies is very sensitive to the galaxy density, the distribution of $W_{0}$(H$_{\alpha}$), the equivalent width of H$\alpha$,  in the star-forming galaxies is independent of environment.
\citet{Balogh04b} studied the color distributions of galaxies in different environments including the typical environment of a cluster core.
After dividing their sample into red and blue components in several luminosity bins, they found that the ratio of these two components is a strong function of the local density but the mean value and the shape of the color distribution of each component are nearly independent of environment. 
They proposed that most star-forming galaxies today evolve independently from their environment.
Both these works suggest that interactions of the galaxies, probably happening in a very short time scale ($\tau < 0.5$ Gyr), may be responsible for triggering star formation and leave the galaxies to evolve afterwards independently from their environments.
Our results appear to agree well with these works and push the independence of the global SFR of IR galaxies on environment to an even lower level of SFR.

It is also possible that the lack of difference between the shape of the Coma LF and LFs from the general field results in part from the effects of averaging. 
First, the general field also includes galaxies in clusters.
Secondly, and more importantly, the galaxies in the outskirt region have evolved largely in the field and are only beginning to fall into the cluster. 
However, their LF of dominates the total LF of the cluster at luminosities above $L_{ir}^{*}$.
Since the outskirt LF has a similar $L_{ir}^{*}$ and a little smaller $\alpha$ compared with the total LF, it is possible that the contribution from the dense region mostly goes to the faint end of the LF where we do detect an environmental effect. 
The early-type galaxies also contribute greatly to the number counts in this range.
However, exact comparisons are complicated by the incompleteness. 
It has been reported that there is a group of dwarf galaxies contributing to the steeply rising faint end slope of the optical and near-IR LF in the Coma cluster \citep{Bernstein95, Secker96, Trentham98, Mobasher98,Pro98}.
These dwarf galaxies are beyond our completeness limit ($R < 19$ mag) and are not the contributors of the steeply rising faint end of the IR LF.

Although we did not obtain any direct evidence of a correlation of SFR and environment in the comparison of the shape of the Coma IR LF and the general field IR LF, there is some evidence of a change in SFR with the environment inside the cluster.
We found that although the fractions of the galaxy members detected at 24 \micron\ do not change very much across the cluster, the LF in the core region has a flatter faint end and is lacking the bright IR galaxies compared to the LF in the outer region.  
In addition, all the galaxies with $L_{ir}^{*} > 10^{44}~{\rm ergs~s}^{-1}$ lie outside of the core region.
This behavior shows that the strongest star-forming activity happens in the lower density region of the cluster.
This is consistent with the speculation that galaxies in a crowded environment lack gas and dust due to the interactions between the galaxies or between the galaxies and the cluster potential well and therefore can not support a large SFR. 
The flatter faint end in the core region, on the other hand, suggests a deficiency of faint IR galaxies as well.

\citet{Mori00} studied the gas stripping of dwarf galaxies by ram pressure in the cluster and found that dwarf galaxies will lose virtually all of their gas instantaneously if their core mass is smaller than a critical mass ($M_{cr}$).
For the Coma cluster, log$(M_{cr}/M_{\sun}$) is about 10.9 at the median distance of the cluster galaxies.
This value increases to about 11.7 and 12.9 at $r=24\arcmin$ ($\sim$ 0.6 Mpc) and $r=12\arcmin$ ($\sim$ 0.3 Mpc).
\citet{Mori00} would argue that galaxies with core masses smaller than these values will lose all of their gas very quickly ($\tau \approx 10^{8}~{\rm yr}$).
Thus, even if the triggering of the star-forming activities in these galaxies happens in the galaxy group before they fall into the cluster core, gas stripping will prohibit them from keeping up such activities by depriving them of fuel.  

However, such an effective stripping contradicts our result.
The faint IR galaxies in our LFs ($41.9 < {\rm log}L_{ir} < 42.6$, about the range of the lowest two points in the LFs above the incompleteness) have $R \approx 15$ mag on average.
The early type galaxies in the Coma cluster usually have a stellar mass-to-light ratio smaller than 8 \citep{Jorgensen99}.
The stellar mass is comparable to the core mass of a galaxy, so we can use this ratio to estimate the core mass of a galaxy from its luminosity.
Considering that faint galaxies usually have relatively large M/L, we take M/L $\approx 20$ as a conservative upper limit for these galaxies.
We also assume all cluster galaxies have similar M/L ratios.
With this ratio and the average $R$ magnitude of the faint IR galaxies, we obtained the upper limit of the mass for these galaxies as log$(M/M_{\sun})\approx 11.0$.
This mass is smaller than the critical mass for gas stripping at $r<24\arcmin$ ($\sim$ 0.6 Mpc), which means most of the faint galaxies in this region should already have lost all of their gas, so we would expect a drop in the number counts of the IR LF.
However, the IR LF in the annulus region does not show such a drop, although we do see a flattening of the slope in the core region.
If this flattening of the faint end slope is the result of the total gas stripping, there seems to be a factor of ten discrepancy between the critical mass Mori \& Burkert deduced and the one suggested by the IR LFs.
A smaller critical mass for these IR galaxies is also consistent with the small difference we found in the Coma IR LF and the IR LF of general field. 
\citet{Mori00} also pointed out that there were some issues they did not consider in their simulation that may affect the critical mass, e.g., the heating of the gas from star forming activities.
For the IR galaxies, this effect may be very important because it would lower the critical mass and provide a possible explanation of why these faint galaxies still have star forming activities.
An underestimated M/L ratio can also contribute to this difference.
However, to resolve the discrepancy in this way, the M/L would need to be as high as 100, which seems unlikely.

We also observed a high ratio of early-type galaxies in the core region of the Coma cluster, $\sim$ 80\%; this ratio drops to about 50\% for the other regions.
The change of the shape of IR LF in the core region is also possibly caused by a morphology-environment correlation rather than a SFR-environment correlation.

\section{Conclusions}
Using MIPS 24 \micron\ observations and two spectroscopic surveys of the Coma cluster, we present the IR LF of the cluster.
The shape of the Coma cluster LF does not differ from that of the general field significantly.
The $L^{*}_{ir}$ value of our LF is very similar to those given by \citet{Rush93} and \citet{Pablo05}, which are both based on surveys of general fields.
The faint end slope of the Coma cluster is shallower than the slope of \citet{Rush93} but steeper than that of \citet{Pablo05} and \citet{Takeuchi03}, again indicating little variation between field and cluster. 
In addition, the overall proportion of IR-active galaxies in the cluster is only slightly less than in the field.
Thus, the overall pattern of star formation in cluster members is surprisingly similar to that in the field, despite an increased galaxy space density by an average factor of $\sim$ 40.

However, in the cluster core where the galaxy density is six times higher still, we found a shallower faint end slope and a smaller $L^{*}_{ir}$ 
compared to the outer region of the cluster, which indicates a decrease in the number of faint IR galaxies as well as in the very bright ones.
The IR-bright galaxies are distributed around the outer region of the cluster.
All the galaxies with $L_{ir} > 10^{44}~{\rm ergs~s}^{-1}$ lie outside of the core region, e.g, $r > 340$ kpc.
No special feature of the IR LF was found in the NGC 4839 region.

In determining the LF of different morphological types, we found that early type galaxies only make about a 15\% contribution to the total IR luminosity density, but they dominate the number density at the low luminosity end.
The global SFR density in the cluster is about 0.88 $M_{\sun}{\rm yr}^{-1}$Mpc$^{-3}$ and the total SFR in the 8.5 Mpc$^{2}$ area of the central cluster is about 97.0 $M_{\sun}{\rm yr}^{-1}$.

\acknowledgments
This work was supported by funding for \textit{Spitzer} GTO programs by NASA, 
through the Jet Propulsion Laboratory subcontracts \#960785 and \#1256318.
We thank Emeric LeFloc'h, Casey Papovich, Pablo P\'{e}rez-Gonz\'{a}lez and Kim-Vy H. Tran for helpful discussions.
\clearpage

\clearpage
\begin{deluxetable}{lrr}
\tablecolumns{3}
\tablecaption{The best fitting parameters of the LFs in different regions.}
\tablehead{Region& log$(L_{ir}^{*}/L_{\sun})$ & $\alpha$\\}
\startdata
NGC 4839 $r < 12'$ &$10.21^{+1.27}_{-0.48}$ &$1.34^{+0.30}_{-0.33}$\\ 
core $r < 12'$ &$9.44^{+0.27}_{-0.37}$ & $0.99^{+0.34}_{-0.52}$\\
 $12' < r < 24'$ & $10.68^{+1.26}_{-0.51}$ & $1.42^{+0.20}_{-0.20}$\\
outskirt & $10.50^{+0.40}_{-0.36}$ & $1.32^{+0.11}_{-0.14}$\\
\enddata
\end{deluxetable}

\end{document}